\documentstyle[preprint,pra,aps]{revtex}
\begin{document}
\draft
\title{\bf
Anomalous anapole moment of an exotic nucleus
}
\author{M.S. Hussein$^{1}$, A.F.R. de Toledo Piza$^{1}$, 
O.K. Vorov$^{1}$,
and A.K. Kerman$^2$ }
\address{
$^{1}$ Instituto de Fisica, 
Universidade de Sao Paulo ,  
Sao Paulo, SP, Brasil \\  
$^2$ Center for Theoretical Physics, \\
Laboratory for Nuclear Science and Department of Physics,\\
Massachusetts Institute of Technology,\\
Cambridge, Massachusetts 02139, USA
}
\date{15 October 1999}
\maketitle
\begin{abstract}
Using the information on the nuclear structure of 
exotic neutron-rich 
halo
nucleus $^{11}$Be, we evaluate the
parity violating anapole moment in its ground state.
The resulting value $\kappa(^{11}$Be)$=0.17$ is fifteen times
bigger than the typical value of the anapole moment
of a normal nucleus of the same mass, 
and in fact exceeds by few times anapole moments 
of any known neutron-odd nuclei
(e.g., $\kappa(^{11}$Be) $>$ $2|\kappa(^{207}$Pb)$|$.
It is also few times bigger than the neutral current 
contribution to the lepton-nucleus interaction.
\end{abstract}

The nuclear anapole moment is one of the most interesting
manifestations 
\cite{ZELDOVICH,FK,FKS84-FKS86,HHM,BP,DKT,EXP,Haxton-SC,EXP-AM-SC}
of the spatial Parity Nonconservation (PNC) \cite{DDH,ADELBERGER}
in atomic physics.
It arises from the PNC nuclear forces which create 
anomalous (toroidal) contribution to the electromagnetic current.
The resulting PNC magnetic field can be experienced by 
an external lepton 
(e.g., atomic electron or muon in mesic atom)
and can be detected in hyperfine structure
atomic measurements.
There were several studies of the nuclear anapole moments 
\cite{FK,FKS84-FKS86,HHM,BP,DKT}.
The first calculation of the quantity in the 
single-particle approximation has been done
by Flambaum, Khriplovich and Sushkov in Ref.\cite{FK}.
Calculation of the  
anapole moment 
with accounting for residual interaction
has been
made by Haxton, Henley and Musolf \cite{HHM}.
Recently, in Refs.
\cite{DKT}
various many-body corrections to the anapole moments
(basically, the many-body contributions to the current)
have been taken into account.
At present, this field attracts much attention 
\cite{DKT,EXP,Haxton-SC,EXP-AM-SC,WB}
as the first experimental results for
the nuclear anapole in $Cs$ are already available \cite{EXP},\cite{EXP-AM-SC}.

Up to now, only the anapole moments of 
normal
nuclei have been 
considered. 
The physics of ``exotic'' nuclei studied with radioactive 
nuclear beams 
\cite{JO-ALK-TO,SAGAWA,HALO-OBSOR,Hencken-Bertsch-Esbensen,PAPER-RESTORATION,Brown-Hansen,Karataglidis-Bennhold,Ren-Faessler-Bobyk,CORE-EXC,NAZ,MOWH,Jnov,BE11-NEW}
appears to be one of the most 
promising
modern nuclear areas.
Specific structure of exotic nuclei
can 
offer
new possibilities to probe those aspects of nuclear interactions 
which are not accessible with normal nuclei.  
The problem
of the PNC effects
in exotic nuclei has been addressed only recently in Ref. \cite{HPVK-HALO} 
where it was shown that
the PNC mixing in halo nuclei can be considerably
enhanced
as compared to the case of ``normal'' nuclei.
It is therefore interesting to examine the anapole moments
of 
exotic nuclei.

Here, we evaluate the 
anapole moment of  
an exotic halo nucleus,
focusing on
the case of $^{11}Be$ which has been extensively studied
both experimentally and theoretically 
\cite{JO-ALK-TO,SAGAWA,HALO-OBSOR,Hencken-Bertsch-Esbensen,PAPER-RESTORATION}. 
We call the resulting anapole moment ``anomalous''
as it exceeds by fifteen times the average anapole
moment of a normal nucleus of the same mass
and is bigger than the anapole moment of any known 
neutron-odd nucleus.
The value of the anapole moment is even twice 
bigger than that of lead.

The Hamiltonian of the nucleus-lepton system can be written
in the form
\begin{equation}
\label{H}
H=H^n_{0} \quad + \quad V^n_{res} \quad + \quad W^n_{PNC} \quad + 
\quad H^{n-e}_{PNC}+ 
\quad h^{n-e}_{PNC},
\end{equation}
where 
$
H^n_{0}=\sum_{i} \left[ {\bf p}_i^{2}/2m+U_{S}(r_i) \right]
$
is the single particle 
Hamiltonian of the nucleons with momentum $\vec{p}$ 
and mass $m$ in 
the single-particle potential $U_{S}(r)$; 
$V^n_{res}$ is the residual strong interaction.
The operator 
$W^n_{PNC}$ is the weak PNC nucleon-nucleon
interaction \cite{DDH}.
The term $H^{n-e}_{PNC}$ describes the interaction of the lepton
with the vector potential $\vec{A}_{PNC}$ created by the nucleus, in which 
we save only the PNC part,
\begin{equation} \label{H-EL}
H^{n-e}_{PNC} = e ( \vec{\alpha} \vec{A}_{PNC} ) = 
e ( \vec{\alpha} \langle \vec{a} \rangle )
\Delta(\vec{r})
\end{equation}
where $\vec{\alpha}$ denote the Dirac matrices 
\cite{BM}
for the lepton
and $\Delta(\vec{r})$ is a function sharply peaked in the
region of the nucleus, it reduces to the $\delta$-function on 
the scale of the atomic electron spatial motion,
$e$ is the proton charge, 
$e^2=\frac{1}{137}$. 
The last term, $h^{n-e}_{PNC}$, is the part of the
neutral current interaction contributing to
the PNC nucleus-lepton forces depending on nuclear spin,
\begin{equation}
h^{n-e}_{PNC} = 
\kappa_{nc} \quad \frac{G}{\sqrt{2}} 
\frac{[ 1/2 - (-1)^{j+l+1/2}(j+1/2)]}{j(j+1)}
( \vec{j} \vec{\alpha} )\Delta(\vec{r}),
\end{equation}
where $\kappa_{nc}\equiv (5/8)(1-4sin^2\theta)$ with
$\theta$ the Weinberg angle.
The vector
$\langle\vec{a}\rangle$ 
is
the expectation value of the anapole moment operator
\begin{equation} \label{ANAPOLE-CURRENT} 
\vec{a}=-\pi\int d^3r r^2 \vec{J}
\end{equation}
in the nuclear
ground state, where  $\vec{J}$ is the nuclear electromagnetic current.
Its is convenient to define the ``anapole moment'',
$\kappa$, rewriting Eq.(\ref{H-EL}) according to \cite{FKS84-FKS86}
\begin{equation} \label{DEFINITION-KAPPA}
H^{n-e}_{PNC} = e ( \vec{\alpha} \langle \vec{a} \rangle ) \Delta(\vec{r})
\equiv
\kappa \quad \frac{G}{\sqrt{2}} \frac{ (-1)^{j+l+1/2}(j+1/2)}{j(j+1)}
( \vec{j} \vec{\alpha} )
\Delta(\vec{r}),
\end{equation}
where $\vec{j}$ is the nuclear 
spin
in the ground
state which coincides with the angular momentum of the external
nucleon if one works in the single-particle approximation;
where $G=10^{-5}m^{-2}$ is the Fermi constant and $m$ is the nucleon mass.
The factors depending on $j$ and on the orbital angular $l$
of the external nucleon absorb the spin-angular dependence of the
anapole expectation value $\langle\vec{a}\rangle$, and 
the anapole moment $\kappa$ chosen in this way contains merely
the nuclear structure information.

In the single-particle approximation, the anapole moment operator
(\ref{ANAPOLE-CURRENT}) is the sum of the spin- and orbital terms
\cite{FKS84-FKS86}
\begin{equation}
\label{COUPLING}
\vec{a}
=  
\frac{\pi e}{m} \sum_i
\left( \mu_i \vec{r}_i \times \vec{\sigma}_i  +  
\frac{q_i}{2} \{\vec{p}_i,r_i^2\} \right).
\end{equation}
where $\vec{\sigma}$ are the spin Pauli matrices,
$\mu$ are the nucleon magnetic moments
[$+2.79$ for proton and $-1.91$ for neutron], $q$ measures 
nucleon charge [$q=1(0)$ for protons (neutrons)] and
$\{,\}$ denotes anticommutator.
Eq.(\ref{COUPLING}) neglects the corrections which come from the 
interactions contributions (e.g., from the weak forces)
to the ectromagnetic current \cite{FK},\cite{DKT}; this is reasonable 
for the simplest estimate, especially, in the case of 
neutron valence nucleon under consideration.
The expectation value of (\ref{COUPLING}) 
in any eigenstate of the nuclear Hamiltonian,
$H^n_0+V^n_{res}$ iz zero
unless parity violating forces 
$W^n_{PNC}$ are taken into account. 
As a result of 
the
PNC weak interaction $W^n_{PNC}$ in the 
Hamiltonian (\ref{H}), a nuclear state of definite parity
$|\psi \rangle$, 
acquires very small 
admixtures of wrong parity configurations
$|{\bar \psi}_n \rangle$. This can be accounted for
by using the first order of perturbation theory with respect 
to $W^n_{PNC}$.
Thus
the expectation value of the anapole moment operator $\vec{a}$
in the state $|{\widetilde\psi} \rangle$ with energy $E$
containing the PNC admixtures is
\begin{equation} \label{ANAPOLE}
\langle \widetilde{\psi} | \vec{a}  |   \widetilde{\psi} \rangle =
\sum_{n}
\left( \frac{ \langle \psi | W_{PNC} | {\bar \psi}_n\rangle }{E-E_{n}}
 \langle {\bar \psi}_n | \vec{a} |\psi \rangle -
 \langle \psi | \vec{a} | {\bar \psi}_n\rangle 
\frac{ \langle {\bar \psi}_n| W_{PNC} | \psi\rangle }{E_{n}-E}
\right)
\end{equation}
where sum runs over the opposite parity states $|{\bar \psi}_n\rangle$.
In a finite nucleus, a nucleon experiences the combined action
of the two-body PNC forces $W_{PNC}$
\cite{DDH}
from other nucleons, 
which
can be modeled \cite{ADELBERGER} by 
the effective one-body PNC weak potential
$w_{PNC}$ 
\begin{eqnarray}  \label{WEAK-POTENTIAL}
w_{PNC}= 
g\frac{G}{2\sqrt{2}m}
\lbrace({\vec \sigma}{\vec p}),\rho\rbrace ,
\end{eqnarray}
The nuclear core density 
$\rho=\sum\limits_{occ} |\psi_{occ}|^2$ 
in (\ref{WEAK-POTENTIAL}) reflects the 
coherent contribution from all the 
occupied nucleon orbitals.
The dimensionless constants $g$ for proton and neutron 
are 
$
g_p = 4.5 \pm 2$,
$g_n = 1 \pm 1.5.
$
These widely used
values 
\cite{ADELBERGER,FK,FKS84-FKS86,we}
correspond to the best values \cite{DDH} of the
microscopic parameters in the DDH
Hamiltonian 
\cite{DDH}.
They are found in reasonable 
agreement with 
the bulk experimental data on 
PNC
including the compound nuclear experiments by TRIPLE group
\cite{te0}
and
anapole moments 
of stable nuclei \cite{EXP-AM-SC}.

The basic specific properties of the halo nuclei are 
determined by the fact 
of existence of loosely bound nucleon in addition to the 
core composed by the rest of the nucleons \cite{HALO-OBSOR}.

In one-body halo
nuclei like $^{11}Be$, the ground state is particularly
simple: it can be 
represented as direct product of the single-particle wave function
of the external neutron,
$\psi_{halo}$, and the wave function of the core.
The spin-saturated 
core does not contribute to
(\ref{ANAPOLE}).
The residual interaction $V^n_{res}$ in (\ref{H})
can be neglected
as the many-body effects related to the core
excitations are generically weak in such nuclei \cite{CORE-EXC}.
As a result of the
relatively heavy core 
for $A\simeq 10$,
difference between 
the center of mass coordinate and the 
center of core coordinate
can also be neglected.  
The problem with the Hamiltonian (\ref{H}) and (\ref{WEAK-POTENTIAL})
is reduced to a 
single-particle problem for the external nucleon.
For the nucleus with the external {\it neutron},
as is the case for the halo nucleus $^{11}$Be, the orbital
part of the anapole operator (\ref{COUPLING}) does not contribute.
Using the 
reduced matrix elements of 
$
\langle l',j,m||\frac{\vec{r}}{r} \times \vec{\sigma}||  l,j,m \rangle =
i (-1)^{j+l+1/2}(j+1/2)\sqrt{\frac{2j+1}{j(j+1)}    },
\quad l'= l \pm 1,
$
the expression for the anapole moment in terms of the radial wave
functions $R_{nlj}$ is
\begin{eqnarray} \label{FINAL-KAPPA}
\kappa = -\frac{2\pi \mu_n e^2 g_n}{m^2}
\sum\limits_{n'l'j}
\frac{   
\int\limits_0^{\infty} r^2dr R_{n'l'j} \left[
\rho \left(
\frac{dR_{nlj}}{dr}+\frac{(l-j)(2l+1-j)}{r}R_{nlj} 
\right)
+\frac{d\rho}{2dr}R_{nlj}
\right]
}
{E_{nlj}-E_{n'l'j}    }
\int\limits_0^{\infty} r^3drR_{n'l'j}R_{nlj}
\end{eqnarray}
In a halo nucleus like $^{11}$Be or $^{11}$Li,
the energy spacing between the opposite parity weakly-bound states
can be small 
\cite{JO-ALK-TO,SAGAWA,HALO-OBSOR,Hencken-Bertsch-Esbensen,PAPER-RESTORATION,CORE-EXC}.
The PNC effect
in (\ref{FINAL-KAPPA}) can therefore be considerably magnified
\cite{HPVK-HALO}.
The nucleus $^{11}Be$ has the only bound excited state, $1p_{1/2}$,
above 
the ground state $2s_{1/2}$ 
\cite{JO-ALK-TO,SAGAWA},
\cite{Hencken-Bertsch-Esbensen},\cite{PAPER-RESTORATION}
(the well known ``inversion of levels''). 
As a result of the 
small energy separation between these levels of
opposite parities 
which is known experimentally,
\begin{equation} \label{DELTA-E}
| \Delta E | = E_{p 1/2} - E_{s 1/2} = 0.32 MeV ,
\end{equation}
one can save the only $1p_{1/2}$ term in the expression (\ref{FINAL-KAPPA})
for the anapole moment $\kappa$ of the ground state $2s_{1/2}$.
The form of the single-particle wave functions 
of halo states can be deduced from their basic properties
\cite{PAPER-RESTORATION} 
and their quantum numbers \cite{SAGAWA}. The results of the 
Hartree-Fock calculations which reproduce the main halo properties
(e.g., mean square radii) are also available \cite{SAGAWA}. 
We use the following ansatz 
\cite{HPVK-HALO},\cite{FHKV}
for the model wave functions of the $2s$ 
and the excited $1p$ 
halo states:
\begin{equation} \label{ANSATZ-2s}
R_{2s}(r) = 
\frac{ 2^{3/2} a^2 (1- (r/a)^2 ) exp(- r/r_0) }
{ r_0^{3/2} \sqrt{45r_0^4 + 2a^4 - 12 a^2r_0^2     } } 
, \quad 
R_{1p}(r) = 
\frac{ 2}{\sqrt{3}} r_1^{-5/2}
exp(- r/r_1),
\end{equation}
The values of the 
parameters $r_0$, $a$ and  
$r_1$ 
must be chosen
to fit the density distributions
\cite{SAGAWA} and the mean square radii
\begin{eqnarray} \label{RADII}
\langle r_{2s}^2\rangle = r_0^2
\frac{6(45r_0^4 + 2a^4 - 12 a^2r_0^2   )}
{105r_0^4 + a^4 - 15 a^2r_0^2 }, 
\langle r_{1p}^2 \rangle = \frac{15}{2}r_1^2.
\end{eqnarray}
The value of $a= 2 fm$  in the wave function of the $2s$ state
is determined by the position of the node 
\cite{SAGAWA},\cite{PAPER-RESTORATION} which
can be extracted, e.g., from the neutron scattering
experiments
\cite{PAPER-RESTORATION}. 
The core nucleon density $\rho_c(r)$ has been 
taken 
according to
Ref. \cite{SAGAWA}
\begin{equation} \label{ANSATZ-CORE}
\rho_c(r) = \rho_0 e^{ -\left(r/ R_c \right)^2},
\rho_0 = 0.2 fm^{-3}, \quad R_c = 2 fm,
\end{equation}
as shown on Fig.1.
Evaluation of (\ref{FINAL-KAPPA}) with the wave functions (\ref{ANSATZ-2s})
and the core density (\ref{ANSATZ-CORE})
gives the expression for the anapole moment in terms of the
parameters:
\begin{eqnarray} \label{INTEGRAL-ANALYT}
\kappa \quad = \quad   
\frac{\pi \mu e^2g_n \rho_0 
}{  m^2 |\Delta E| }  
\quad 
\frac{4 R_c^{10} a^2 \left[ a^2(r_0+r_1)^2-30r_0^2r_1^2    \right]
  }
{ r_0^5 r_1^7 y^7( 45 r_0^4 +2a^4 -12 a^2 r_0^2 )   } \times
\qquad \qquad \qquad 
\nonumber\\  \times
\biggl\{ 3I_2(y) -\left[ 3\left(\frac{R_c}{a}\right)^2+1\right] I_4(y) +
\left(\frac{R_c}{a}\right)^2 I_6(y)
-
\frac{R_c}{r_1}\left[I_3(y)-\left(\frac{R_c}{a}\right)^2 I_5(y)\right]
\biggr\} 
\end{eqnarray}
where 
$y=\frac{R_c(r_0+r_1)}{2 r_0 r_1}$ 
and the functions 
$
I_n(y) = 
(-1)^n \frac{\sqrt{\pi}}{2^{n+1}}  \frac{d^n}{dy^n} e^{y^2} erfc(y),
$
are given 
in terms of 
the error function 
$
erfc(y) = 1-\frac{2}{\sqrt{\pi}} \int\limits_0^y dt \quad exp( -t^2/2).
$

The results for the densities calculated with 
the optimal values of the parameters,
$r_0=1.45$ fm, $r_1=1.80$ fm,
are shown in Fig.1.
One sees good agreement with the Hartree-Fock calculations \cite{SAGAWA}.
The values of the halo radii 
given by (\ref{RADII}),
$\sqrt{\langle r_{2s}^2\rangle}=5.9 fm$ and 
$\sqrt{\langle r_{1p}^2\rangle} = 4.9 fm$ are 
close to 
the values of Ref.\cite{SAGAWA}   $6.5 fm $ and $5.9 fm$
which agree with experimental matter radii.

With the above values of the parameters, 
we finally obtain from (\ref{FINAL-KAPPA})  the resulting value of the 
anapole moment $\kappa$
\begin{eqnarray} \label{W-RESULT-NUMERIC} 
\kappa \left( ^{11} Be \right) =  0.17 g_n =  0.17
\quad ( for  \quad g_n \simeq 1 \quad).
\end{eqnarray}
It is  
few
times bigger than 
the contribution from
neutral current $-\kappa_{nc}=-0.05$, 
thus the nuclear spin-dependent PNC interaction of a lepton
with the halo nucleus
is dominated by the anapole moment
contribution, as in heavy nuclei.

To appreciate how big the value $\kappa\left(^{11}Be\right)$ 
is, one can compare (\ref{W-RESULT-NUMERIC}) to 
the 
anapole moment of the normal spherical nucleus with odd neutron 
which is given by 
\cite{FKS84-FKS86}:
\begin{equation}  \label{NORMAL-KAPPA}
\kappa_{norm} =
\frac{9}{10} \frac{g_n e^2 \mu_n}{m r_0} A^{2/3},
\end{equation}
where $r_0=1.2$fm is the nucleon radius.
Resulting from the PNC
toroidal 
electromagnetic currents, the anapole moment  grows fast 
($\propto A^{2/3}$) as the size of 
the system increased \cite{FK,FKS84-FKS86}. 
For this reason, the anapole moments of {\it normal} light nuclei
give only a small correction to the neutral current lepton-nucleus 
PNC interaction
(see Fig. 2).
From (\ref{W-RESULT-NUMERIC}) and (\ref{NORMAL-KAPPA}),
we find the ratio 
of the anapole moment to its value
in a nucleus with the same A ( enhancement factor):
\begin{equation} \label{R=15}
R_{halo} \quad =  \quad \frac{\kappa(^{11}Be)}{\kappa_{norm}}  \quad 
=  \quad  15 .
\end{equation}
In fact,
the anapole moment (\ref{W-RESULT-NUMERIC})
exceeds few times 
the anapole moments of any known odd nucleus, as seen in Fig.2. 
For example, the $\kappa($$^{11}$Be)  
is two times bigger than 
the anapole moment of nucleus as heavy as
lead \cite{FKS84-FKS86},
\begin{displaymath}
\kappa(^{207}Pb) = -0.08 g_n
\end{displaymath}

The remarkable enhancement factor (\ref{R=15})
in (\ref{W-RESULT-NUMERIC}) 
comes from the two features of the halo structure:
a) enhancement of the PNC mixing in the halo ground state 
[the first factor in Eq.(\ref{R-FACTORS})]
and
b) enhancement of the matrix elements of the anapole operator
in halo states:
\begin{equation} \label{R-FACTORS}
R_{halo} \quad \sim  \quad 
\frac{\omega}{\Delta E} \left( \frac{w_{halo}}{w_{norm}}\right)
\left( \frac{r_{halo}}{r_{norm}}\right),
\end{equation}
where the second factor is the ratio of the halo weak matrix element
$w_{halo}$ (\ref{WEAK-POTENTIAL}) to the normal one, $w_{norm}$,
which is less than unity.
The parity violating effect originates from the weak interaction 
of the external halo
neutron with the core nucleons in the nuclear interior.
As a result, the neutron halo cloud surrounding the nucleus 
acquires the wrong parity admixtures.
Those give rise to the PNC toroidal currents  
in the nuclear
exterior (the halo region) which results in additional enhancement
of the anapole moment [the last factor in (\ref{R-FACTORS})].

We discuss now the stability of the results
against possible distortions of the wave functions (\ref{ANSATZ-2s})
we used.
Table I shows the values of the anapole moment 
calculated for various values of the parameters $r_0$ and $r_1$
in the wave functions (\ref{ANSATZ-2s}).
As is seen from the Table,
the results are stable 
with respect to variation of the details of the halo structure.

We consider now the influence of the many-body contributions 
(see, e.g., \cite{MOWH,Jnov})
to the
halo wave functions (\ref{ANSATZ-2s}) on the present results.
The generalized wave function of the halo ground state, $|s)$, can be written as a sum 
\begin{eqnarray}  \label{mb}
|s) = (1-x_s^2) |s_{sp}\rangle + x_s |S_{mb}\rangle,
\end{eqnarray}
where $|s_{sp}\rangle$ is the purely single-particle
s-state (\ref{ANSATZ-2s}) and $|S_{mb}\rangle$ denotes the many-body
contributions (core polarization) which have not been considered yet.
The coefficient $x_s$ ($0 \leq x_s \leq1$) is the amplitude
of the many-body correction which is properly normalized,
$\langle S_{mb}|S_{mb}\rangle=1$.
The anapole moment can be evaluated in the same way as above,
using Eqs.(\ref{COUPLING}), (\ref{ANAPOLE}) and (\ref{WEAK-POTENTIAL}) 
and the state $|s)$ instead of $|s_{sp}\rangle$.
Both the anapole moment operator (\ref{COUPLING}),
the weak potential (\ref{WEAK-POTENTIAL}) are the single-particle operators,
so they can not connect the single-particle wave function
$|p_{sp}\rangle$ (\ref{ANSATZ-2s})
with the many-body component $|S_{mb}\rangle$, thus
$\langle S_{mb} | \vec{a} |p_{sp}\rangle = 0$ and 
$\langle S_{mb} | w_{PNC} |p_{sp}\rangle = 0$. 
The anapole moment ${\tilde \kappa}$ in the state $|s)$ 
is 
now given by
\begin{displaymath}
{\tilde \kappa} = (1-x_s^2) \kappa,
\end{displaymath}
where $\kappa$ is the single-particle result (\ref{FINAL-KAPPA}),
(\ref{INTEGRAL-ANALYT}) and (\ref{W-RESULT-NUMERIC}).
Similarly to (\ref{mb}), one can consider many-body contributions
$|P_{mb}\rangle$ to 
the excited p-state $|p_{sp}\rangle$ (\ref{ANSATZ-2s})
with the amplitude $x_p$,
$|p) = (1-x_p^2) |p_{sp}\rangle + x_p |P_{mb}\rangle$. In this case,  
the result is
\begin{displaymath}
{\tilde \kappa} = \kappa\left[
(1-x_s^2) (1-x_p^2) + x_s x_p \sqrt{(1-x_s^2) (1-x_p^2)}(u+v)+
x_s x_p u v \right],
\end{displaymath}
where $u=\frac{\langle S_{mb}|\vec{a}|P_{mb}\rangle}
{\langle s_{mb}|\vec{a}|p_{mb}\rangle}$ and 
$v=\frac{\langle S_{mb}|w_{PNC}|P_{mb}\rangle}
{\langle s_{mb}|w_{PNC}|p_{mb}\rangle}$.
The matrix elements between the many-body wave functions
are generally suppressed as compared to those between the 
single-particle states.
According to some recent experimental results, the many-body
contributions
are rather small,
$\approx 16\%$.  
\cite{BE11-NEW}.
Thus the corrections due to the many-body admixtures
in the halo states are about the same order of magnitude as 
the many-body corrections to the operators (\ref{COUPLING}) and
(\ref{WEAK-POTENTIAL}). They can be taken into account in more refined
calculations using detailed information on the wave function
structure.

The curious ``halo anomaly'' is quite interesting in a number of respects.
First, search for sources of enhancement in anapole moments
has been always important from the experimental viewpoint.
Possibilities offered by the normal nuclei are rather limited here.
The most promising case of deformed nuclei, where one can find
close levels of opposite parity near the ground state, does not
offer any enhancement because of the
suppression in the matrix elements of $\vec{a}$ \cite{FKS84-FKS86}.
In this respect, the anomalies in anapole moments of 
exotic nuclei like $^{11}$Be
seem to give unexpected opportunity.
Secondly, the anapole moment of neutron-rich nuclei
is determined by the neutron weak constant 
$g_n$ only. 
Usually, sensitivity of experiments 
to the value of this constant is ``spoiled'' by relatively large 
value of the proton weak constant $g_p$, in Eq.(\ref{WEAK-POTENTIAL}).
The 
large
enhancement of the anapole moment in neutron halo nuclei
provides therefore an unique opportunity to test the
isospin structure 
of the weak potential (\ref{WEAK-POTENTIAL})
which is at present of great 
interest
\cite{WB}.

One should note that the nucleus $^{11}$ Be has a rather
long life-time 
(13.81 $sec$).
This makes therefore possible the atomic measurements of the
hyperfine structure effects
in traps planned for the future ISOL facility where
the anapole moment can be detected.

The work has been supported by FAPESP 
and  
in part by funds
provided by  the U.S.~Department of Energy (D.O.E.) under contract
\#DE-FC02-94ER40818.

\begin{table}
\caption{
Dependence of $\kappa \left( ^{11} Be \right)$ 
on the halo parameters $r_0$ and $r_1$ in Eq.(\ref{INTEGRAL-ANALYT});
the ratios of  $\kappa$
to the result (\ref{W-RESULT-NUMERIC}) are given.
The central entry in the table corresponds to the optimal values
used in (\ref{W-RESULT-NUMERIC}).
It is seen that variations in $r_0$ and $r_1$ do not affect 
(\ref{W-RESULT-NUMERIC})
any considerably.
} 

\vspace{1cm}

\begin{tabular}{cccccc}   
  & $ r_0=1.35 $ & $ r_0=1.40 $ & $r_0=1.45 $  & $ r_0=1.50$ & $ r_0=1.55$ \\ 
\tableline
$ r_1= 1.70 $ &  $1.26 $ &  $1.15 $  & $1.05 $  & $0.96 $ & $0.87 $  \\
$ r_1= 1.75 $ &  $1.23 $ &  $1.12 $  & $1.03 $  & $0.94 $ & $0.85 $  \\
$ r_1= 1.80 $ &  $1.19 $ &  $1.09 $  & $1.00 $  & $0.91 $ & $0.83 $  \\ 
$ r_1= 1.85 $ &  $1.16 $ &  $1.06 $  & $0.97 $  & $0.89 $ & $0.81 $ \\   
$ r_1= 1.90 $ &  $1.12 $ &  $1.03 $  & $0.95 $  & $0.87 $ & $0.80 $  

\end{tabular}

\end{table}


{\large Figure Captions}

\vspace{1cm}
Fig.1. Densities $R^2(r)$ 
for the halo states $2s$ and $1p$ as function of $r$
and the core density $\rho_c(r)$ calculated from
Eqs.(\ref{ANSATZ-2s},\ref{ANSATZ-CORE}) (solid lines).
The Hartree-Fock results for the same quantities \cite{SAGAWA}
are given by the dashed line, the dotted line and the 
dotted-dashed line, respectively.

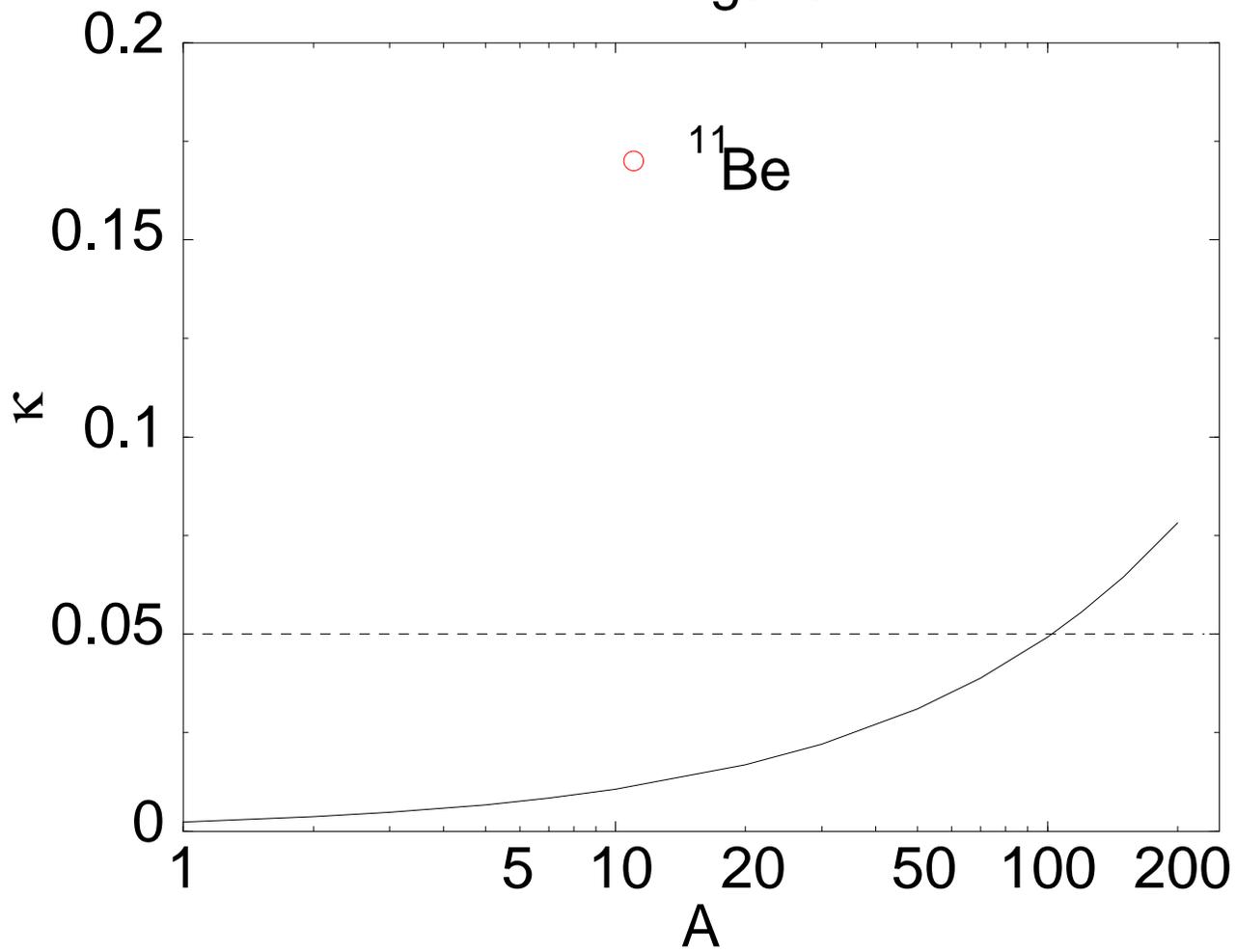
Fig.2. ``Halo anomaly'' in $^{11}$Be: the value 
$\kappa \left( ^{11} Be \right)$ (circle) as compared to 
the 
absolute values of the 
anapole moments of normal neutron-odd nuclei 
(solid curve) and the neutral current contribution
$\kappa_{nc}=0.05$ (dashed curve).

\noindent
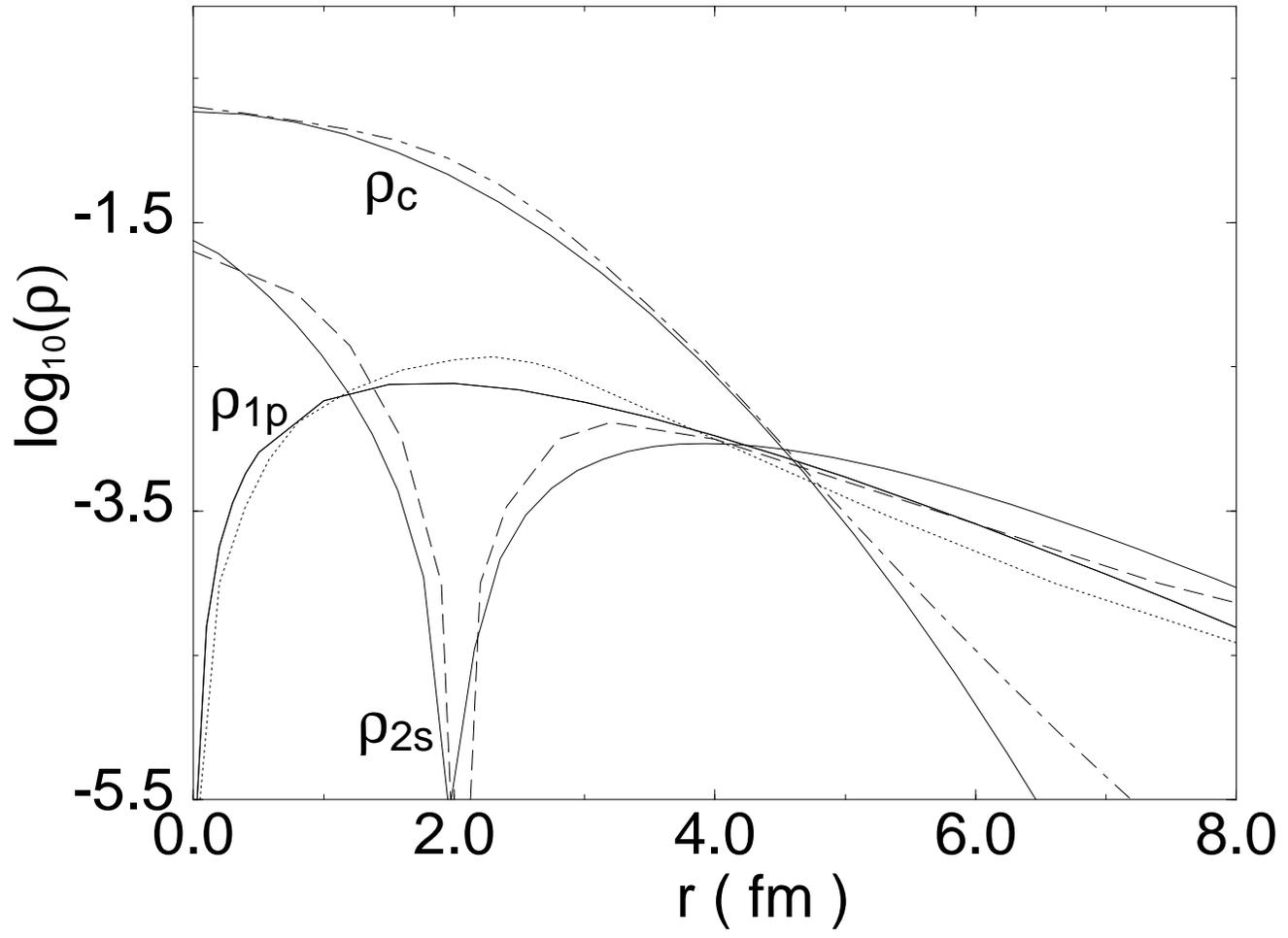
\end{document}